# The distribution of the doped holes in La$_{2-x}$Sr$_x$Cu$_{1-y}$Ru$_y$O$_{4-\delta}$

Z. Hu, M. Knupfer, M. S. Golden, J. Fink

Institute for Solid State Research, IFW Dresden,

P. O. Box 270016, D-01171 Dresden, Germany

S. Ebbinghaus

Lehrstuhl für Festkörperchemie, Institut für Physik, Universität Augsburg

Universitätsstraße 1, 86159 Augsburg, Germany

F.M.F. de Groot

Departement of Inorganic Chemistry and Catalysis, Utrecht University

3584 CA Utrecht, The Netherlands

G. Kaindl

Institut für Experimentalphysik, Freie Universität Berlin

Arnimallee 14, D-14195 Berlin-Dahlem

**Abstract**

We present a systematic study of the perovskite-related system La$_{2-x}$Sr$_x$Cu$_{1-y}$Ru$_y$O$_4$ using Cu-L$_{2,3}$, Ru-L$_{2,3}$, and O-K x-ray absorption spectroscopy (XAS). This system can be regarded as a solid solution of the two superconductors La$_{2-x}$Sr$_x$CuO$_4$ and Sr$_2$RuO$_4$, and thus the question as to the destination of the holes induced by Sr doping is central to our understanding of these novel systems. The comparison of the experimental data with state-of-the-art crystal-field-multiplet calculations shows clearly that the charge balance for $x/2y >1$ is predominantly achieved by an increase of the Ru valence from Ru(IV) to Ru(V), while Cu remains in the Cu(II) oxidation state.

PACS numbers: 78.70. Dm, 71.28.+d, 79.60

## I. Introduction

The discovery of high-temperature superconductors [1] has initiated a large number of studies on the system $La_{2-x}Sr_xCuO_4$, which has the simplest crystal structure among the variety of high-temperature superconductors known today. The parent, unsubstituted compound $La_2CuO_4$ is an antiferromagnetic insulator. Upon Sr-substitution it becomes superconducting with a maximum value of $T_C$ for $x = 0.15$ [2]. At higher doping superconductivity disappears again and $La_{2-x}Sr_xCuO_4$ shows a normal metallic behavior [2]. This has lead to further extensive research of the consequences of substituting the La or Cu atoms by other metals.

Recently, superconductivity was also found in the related system $Sr_2RuO_4$ at rather low temperatures ($T_C \approx 1$ K) [3]. Very little is known about other members of the $La_{2-x}Sr_xCu_{1-y}Ru_yO_{4-\delta}$ system [4-6], which may be considered to be a solid solution of the two superconductors $Sr_2RuO_4$ and $La_{2-x}Sr_xCuO_4$. Two previous studies [7,8] indicate that single phase $La_{2-x}Sr_xCu_{1-y}Ru_yO_{4-\delta}$ exists in the regions $0 \leq x \leq 2$ and $0 \leq y \leq 1$ with $x \geq 2y$ and $x < 1+2y$, which are illustrated in the schematic phase diagram of Fig. 1. Two of the most important parameters that control the electronic behavior of these materials are the valences of the Cu and Ru ions. For $\delta = 0$ and $x = 2y$, Cu and Ru have the valence states of Cu(II) and of Ru(IV), as in $La_2CuO_4$ and $Sr_2RuO_4$, respectively. For $x > 2y$, thermogravimetric measurements have shown that up to 40% of the Ru or Cu cations adopt a higher oxidation state [7,8], which then is either due to formation of Cu(III) or Ru(V) ions. In the case of high values of the ratio x/2y, the system loses oxygen. This leads finally to a breakdown of the $K_2NiF_4$-type structure when the oxygen deficiency $\delta \leq 0.3$ [7].

The question remains whether for $x > 2y$ charge balance is achieved by hole creation in Cu-3d states, giving rise to Cu(III) ions, or in Ru-4d states resulting in Ru(V). X-ray absorption spectroscopic (XAS) studies at the Cu-$L_{3,2}$ and O-K edges are now standard methods for studying Cu valence and the covalence between Cu-3d and O-2p states in cuprates [9-14]. In addition, our previous combined theoretical and experimental work on model Ru(IV) and Ru(V) oxides indicated that the Ru-$L_{2,3}$ XAS spectra, as concerns both energy position and spectral profile, are sensitive to the valence state of Ru [15]. In this contribution we present a systematic study of the $La_{2-x}Sr_xCu_{1-y}Ru_yO_{4-\delta}$ system using XAS at the Cu-$2p_{1/2,3/2}$ ($L_{2,3}$), Ru-$2p_{1/2,3/2}$ ($L_{2,3}$), and O-1s (K) thresholds with the aim to answer the above questions.

## II.    Experimental

Polycrystalline samples of $La_{2-x}Sr_xCu_{1-x}Ru_xO_{4-\delta}$ were prepared according to the procedures given in Ref. [7], and the $Sr_4Ru_2O_9$ sample was prepared as described in Ref. [16]. The purity of the samples was checked by x-ray diffraction. The magnetic investigations reported here were performed using a Biomagnetic Materials SQUID model VTS-905. The measurements were carried out in the temperature range 6-300 K with an external magnetic field of 0.1 T. Electrical resistivity measurements were performed on pressed powder pellets in the temperature range 300-575 K using a standard four-probe technique. A Hewlett Packard 4284A supply was used to generate a 1 kHz ac-voltage of 10 m V.

The O-K and Cu-$L_{2,3}$ XAS spectra were recorded at the SX700/II monochromator operated by the Freie Universität Berlin at the Berliner Elektronenspeicherring für Synchrotronstrahlung (BESSY) in a fluorescence yield mode. At the O-K and Cu-$L_3$ thresholds, the experimental resolutions were 0.3 eV and 0.6 eV, respectively. The O-K XAS spectra were corrected for the energy-dependent incident photon flux and were normalized 60 eV above threshold. Prior to the measurements, the sample surfaces were scraped *in-situ* with a diamond file at a base pressure of $5*10^{-10}$ mbar. The Ru-$L_{2,3}$ XAS spectra were recorded in transmission geometry at the EXAFS-II beamline at HASYLAB, using a Si(111) double-crystal monochromator. This resulted in an experimental resolution of $\cong 1.2$ eV (FWHM) at the Ru-$L_3$ threshold (2838 eV).

Throughout this contribution, we will refer to the systems in question using capital letters which represent a set of values for x, y and $\delta$ denoted $(x,y,\delta)$ which are summarized in Table I.

## III. Results

### 1. Magnetism and conductivity

A large variation of the magnetic properties was found in the system $La_{2-x}Sr_xCu_{1-y}Ru_yO_{4-\delta}$. In this paper, only a few typical examples are shown. A more extended study will be published elsewhere. As can be seen from Fig. 2, the sample P(1.36,0.2,0.28) (dashed curve) shows an almost ideal Curie-Weiss behavior in the magnetic susceptibility between 6 K and room temperature, which represents the magnetic behavior found for most of the samples studied. In a number of samples, e.g. E(1.64, 0.7,0) (Fig. 2, solid curve), the magnetic susceptibility $\chi_{mol}$ shows a markedly different behavior below 20 K, depending on whether the sample is field-cooled or zero-field-cooled. The data indicate a typical "spin-glass" behavior which had been observed previously by Kim *et al.* for the series $LaSr_nCuRuO_{n+5}$ (with n = 1,2,3) [6]. The third class of magnetic behavior is shown for the sample G(0.74,0.3,0) (Fig. 2, dotted curve), which shows an antiferromagnetic phase transition at $T_N$ ≈ 117 K. Below $T_N$ the field-cooled and the zero-field-cooled curves are only slightly split. This class of compounds was found to possess an unusually large a axis and a short c axis [7].

The electrical resistivity ρ exhibits a maximum at y ≈ 0.5 for a given x, in agreement with previous work [6], where an increase of ρ with y in the region y < 0.5 had been reported. This could be a consequence of the disorder induced by the randomly distributed Cu and Ru atoms in the $(Cu,Ru)O_2$ planes.

*2. Cu-$L_3$ XAS spectra*

Figure 3 shows the Cu-$L_{2,3}$ XAS spectra of some selected compounds of the series $La_{2-x}Sr_xCu_{1-y}Ru_yO_{4-\delta}$, together with those of CuO and $La_2Li_{0.5}Cu_{0.5}O_4$ as Cu(II) and Cu(III) references, respectively. The strong single peak observed for all $La_{2-x}Sr_xCu_{1-y}Ru_yO_{4-\delta}$ systems lies at energies ranging from 930.7 eV to 931.4 eV, it is shifted by only $|\Delta E| \leq 0.5$ eV compared to that of CuO at 931.2 eV, the latter structure being due to a $\underline{2p}3d^{10}$ final state arising from the $3d^9$ initial state of Cu(II). In contrast, the strong peak in the Cu(III) reference compound $La_2Li_{0.5}Cu_{0.5}O_4$ lies 1.7 eV above the peak of CuO and is attributed to a predominantly $\underline{2p}3d^{10}\underline{L}$ final state ($\underline{L}$ denotes a hole in the O-2p ligand orbitals). The weak satellite at ~ 9 eV above the main peak in $La_2Li_{0.5}Cu_{0.5}O_4$ is assigned to a predominantly $\underline{2p}3d^9$ final state, which exhibits a multiplet splitting [9,14,17]. The Cu-$L_{2,3}$ edges reveal that in most compounds discussed here the copper ions remain divalent [Cu(II)]. Only at very high x/2y ratios (close to the breakdown of the $K_2NiF_4$-type structure), e.g. for P(1.36;0.2,0.28), can one see a weak shoulder (indicated by an arrow in Fig. 3) at the energy of the main peak of $La_2Li_{0.5}Cu_{0.5}O_4$. Such a weak shoulder is also observed previously in $La_{2-x}Sr_xCuO_4$ [10,11,18], where the holes created by Sr doping have mainly O-2p, but essentially no Cu-3d character.

*3. Ru-$L_{2,3}$ XAS spectra*

The unchanged Cu valence leads to the conclusion that the Ru ions play the central role as regards the charge balance in these compounds. It is usually assumed that the $L_{2,3}$ XAS spectra of 4d TM compounds reflect directly unoccupied 4d orbitals, and the spectra have correspondingly often been interpreted in terms of crystal-field or molecular-orbital theories [19-22]. In the case of $O_h$ local symmetry, it is then expected that the intensity ratio of transitions into the crystal-field-split $t_{2g}$ and $e_g$ states, ($I(t_{2g})/I(e_g)$, is 0.5 for Ru(IV)($4d^4$) and increases to 0.75 for Ru(V)($4d^3$) at both the $L_2$ and the $L_3$ edges. Recently, however, it was recognized that the intra-atomic Coulomb interaction and the 4d spin-orbit coupling strongly modify the spectral features, as concerns both their energy position and intensity ratio of $t_{2g}$- and $e_g$-related peaks in Ru-$L_{2,3}$ XAS spectra [15,23]. Therefore only a combined experimental and theoretical study can give reliable information on the valence of Ru.

For simplicity, we first show in Fig. 4 the Ru-$L_{2,3}$ XAS spectra of $Sr_2Cu_{1-y}Ru_yO_4$ with y = 1[A(2,1,0)], 0.9[C(2,0.9,0)], and 0.7[J(2,0.7,01)] together with those of $Sr_4Ru_2O_9$ (bottom) as a Ru(V) reference. To ease comparison, the $L_2$ spectra (open symbols) have been shifted by 129 eV (Ru2p spin-orbit splitting) and have been multiplied by 2.1 so that the high energy feature matches that of the corresponding feature in the $L_3$ spectra (filled symbols). In each case, the lower and the higher energy component basically reflects transitions into $t_{2g}$ and $e_g$-related states, respectively. The intensity ratio, $I(t_{2g})/I(e_g)$, is higher at the $L_3$ edge than at the $L_2$ edge for Ru(IV) compound $Sr_2RuO_4$, but vice versa for the Ru(V) compound $Sr_4Ru_2O_9$. The observed spectral ratios are significantly different from those expected from crystal-field and molecular-orbital theories, as was discussed in detail in Ref. [15]. In addition, on going from Ru(IV) to Ru(V), the energy position of both components of the $L_2$ and $L_3$ XAS spectra are shifted by 1.5 eV to higher energies.

The changes in the Ru-$L_{2,3}$ XAS spectra described above with increasing Ru valence can be well reproduced by crystal-field-multiplet calculations (CFMC) [15] as shown in Fig. 4. The solid and dashed lines below the spectra of $Sr_2RuO_4$ and $Sr_4Ru_2O_9$ are the theoretical curves obtained for the $L_3$ and $L_2$ edges, respectively. The intensity ratio $I(t_{2g})/I(e_g)$ at both the $L_2$ and $L_3$ edges has been found to be very sensitive to the intra-atomic Coulomb interactions, represented by the corresponding Slater integral and to the 4d spin-orbit coupling [15,22]. Considering the strong covalency between the transition metal d and the O2p states in these systems, the Slater integrals have to be reduced to 50% and 15% of their atomic values for Ru(IV) and Ru(V), respectively [15]. The 4d spin-orbit coupling is also very important. Neglecting it, the $t_{2g}$-related peak for the Ru(IV) $4d^4$ configuration ($Sr_2RuO_4$) should disappear [15], and the intensity ratio $I(t_{2g})/I(e_g)$ should be stronger at $L_3$ than at $L_2$ for the Ru(V) system, $Sr_4Ru_2O_9$, which is plainly not the case. On the basis of the experimental and theoretical spectra for the simple, La-free Ru(IV) and Ru(V) systems, we can conclude that the upward shift in energy positions and the significantly larger $I(t_{2g})/I(e_g)$ ratio at the $L_2$ compared to the $L_3$ edge indicate an increase of the Ru valence with increasing Cu concentration in the $Sr_2Cu_{1-y}Ru_yO_4$ system. We now move on to the more complicated $La_{2-x}Sr_xCu_{1-y}Ru_yO_{4-\delta}$ systems.

The thermogravimetric measurements described in Ref. [7] shows that in $La_{2-x}Sr_xCu_{1-y}Ru_yO_{4-\delta}$ the oxygen deficiency $\delta$ can be approximated by

$$\delta \approx 0 \qquad\qquad \text{for } x \leq 2y + 0.4 \qquad\qquad (1)$$

$$\delta \approx x/2 - y - 0.2 \qquad\qquad \text{for } x > 2y + 0.4. \qquad\qquad (2)$$

Using this approximation, and keeping in mind that Cu remains divalent, the ratio $n^V =$ $N(Ru^V)/[N(Ru^V)+N(Ru^{IV})]$, where $N(Ru^V)$ and $N(Ru^{IV})$ are number of Ru(V) and Ru(IV) ions, respectively, is obtained by

$$n^V = x/y - 2, \qquad \text{for } x \leq 2y + 0.4 \qquad (3)$$
$$n^V = 0.4/y \qquad \text{for } x > 2y + 0.4. \qquad (4)$$

The Ru valence $V$(Ru) is given by

$$V(Ru) = 4 + n^V. \qquad (5)$$

The values of $V$(Ru) obtained according to Eq. (3)-Eq. (5) are summarized in Table 1.

Now we can go back to the phase diagram shown in Fig. 1. The thermogravimetric data indicate that there are three regions for $x > 2y$. The lines $x = 2y$, $x = 3y$ and $x = 2y + 0.4$ make region (1), in which we have the Ru valence $V$(Ru) < 5 and $\delta = 0$. The lines $x = 2y + 0.4$ and $y = 0.4$ make region (2), where we have $V$(Ru) $\leq 5$ and $\delta \geq 0$. Region (3) lies below line $x = 3y$ and $y = 0.4$, where we have $V$(Ru) = 5 and $\delta > 0$. The Ru valences of 4.22 for C(2,0.1,0) [region (1)] and 4.57 for J(2,0.3,0.1) [region (2)] are consistent with the conclusion presented here that the Ru valence increases with increasing Cu concentration in the $Sr_2Cu_{1-y}Ru_yO_{4-\delta}$ system.

The Ru-$L_{2,3}$ spectra of the more complicated $La_{2-x}Sr_xCu_{1-y}Ru_yO_{4-\delta}$ compounds ($x \neq 0$ and $y \neq 0$) are shown in Fig. 5 with increasing the $V$(Ru) from top to bottom. One can see clearly a shift of the peak position to the higher energy and an increase of an intensity ratios of $I(t_{2g})/I(e_g)$ at the Ru-$L_2$ edge with an increase of Ru valence. In the case of the sample N(1.28,0.4,0.04), the Ru valence is Ru(V) i.e. $n^V$ = 1, with the spectral features [both the energy position and the intensity ratios of $I(t_{2g})/I(e_g)$] being the same as those of the Ru(V) reference compound $Sr_4Ru_2O_9$. In the foregoing, we have observed that the intensity ratio $I(t_{2g})/I(e_g)$ in the $L_2$ XAS spectra is the most sensitive spectral parameter to monitor an increase in Ru valence from Ru(IV) to Ru(V). This is a result of the larger intensity transfer from the $e_g$-related to the $t_{2g}$-related peak in the Ru-$L_2$ XAS spectrum (as compared to Ru-$L_3$), which is due to the differences in the intra-atomic interactions in the Ru(IV) [$4d^4$] and Ru(V) [$4d^3$] configurations, as shown in the theoretical spectra of Fig. 4. We stress that this spectral behavior is unexpected from crystal-field or molecular-orbital theories of the Ru-$L_{2,3}$ XAS, which predict the same ratio for the $L_2$ and $L_3$ spectra.

According the rules set up from the thermogravimetric measurements, sample P(1.36,0.2,0.28) lies in region (3) of the phase diagram shown in Fig. 1, and thus $V$(Ru) should be greater than five. However, the experimental Ru-$L_{2,3}$ spectrum is identical with with that of sample N(1.28,0.4,0.04), which has $V$(Ru)= 5 being at the border of regions (2) and (3) in the phase diagram. Inspection of Fig. 5 shows that sample N (and thus also P) has a Ru-$L_{2,3}$ spectrum essentially identical to that of the simple Ru(V) reference compound $Sr_4Ru_2O_9$. In this case of sample P (formally greater than pentavalent), the charge balance is in fact achieved by a large oxygen deficit of $\delta = 0.28$ and by the small number of holes located in the $CuO_6$ units, and seen in the spectra as a shoulder structure in the Cu-$L_{2,3}$ XAS spectrum shown in Fig. 3.

In both Fig. 4 and Fig. 5 one can see a narrowing of the both $t_{2g}$- and $e_g$-related peaks in the Ru-$L_{2,3}$ edges as the Ru valence increases. The reason for this might be that Cu(II) and Ru(IV) are Jahn-Teller ions, causing distortions of the metal-oxygen octahedra of 6-7% and of 27% in $Sr_2RuO_4$ [23] and $La_2CuO_4$ [24], respectively. In $La_{2-x}Sr_xCu_{1-y}Ru_yO_{4-\delta}$ a linear decrease of this distortion with increasing Ru valence is expected, since Ru(V) ions exhibit no Jahn-Teller activity.

In Fig. 6 we summarize the energy shifts of the $t_{2g}$-(filled circles) and $e_g$-related (closed diamonds) features obtained from the Ru-$L_3$ XAS spectra and the intensity ratio $I(t_{2g})/I(e_g)$ (filled squares) from the Ru-$L_2$ spectra as a function of the $V$(Ru). The open symbols show the values for the Ru(V) reference compound $Sr_4Ru_2O_9$.

*4. O-K XAS spectra*

In general, in O-K XAS spectra, the effects of intra-atomic Coulomb correlation are much weaker than in the TM-$L_{2,3}$ XAS spectra, i.e. an agreement between the experimental spectra and the results of band structure calculations is plausible [25-27]. Therefore, O-K XAS spectra are usually studied in order to explore the amount of O-2p holes induced by covalence in the ground state. The O-K XAS spectra of $La_{2-x}Sr_xCu_{1-y}Ru_yO_{4-\delta}$ are rather complicated due to fact that unoccupied O-2p states which result from the covalent hybridization between the Cu-3d and O-2p levels occur at the same energy as those from Ru-4d and O-2p hybridisation. In the following, we discuss the O-K spectra of $Sr_2Cu_{1-y}Ru_yO_4$, together with those of $Sr_4Ru_2O_9$ and CuO which are shown in Fig. 7.

Firstly, we mention that the pre-edge peak at 530.2 eV in the spectrum of CuO, shown at the bottom of Fig. 7, is usually referred to as the upper Hubbard band (UHB), and its intensity reflects the O-2p hole fraction in the predominantly Cu $3d_{x^2-y^2}$-derived UHB due to Cu-3d/O-2p covalence. For either Ru(IV) or Ru(V) in a low-spin state, besides four holes in the $e_g$ orbitals, there are two or three holes, respectively, in the $4d(t_{2g})$ orbitals. Consequentyl, one would expect two pre-edge peaks in the O-K spectra of these systems corresponding to the hybridization footprint of the unoccupied $4d(t_{2g})$ and $4d(e_g)$ states, respectively.

Starting with the Ru(IV) system $Sr_2RuO_4$ (sample A), one sees that whilst the $t_{2g}$-related peak is clearly visible in Fig. 7 at 529.5, the $e_g$-related feature overlaps strongly with Sr4d-related spectral weight, as discussed previously [15]. The same holds for sample C. With increasing Cu concentration in $Sr_2Cu_{1-y}Ru_yO_4$, the Ru valence increases, as was proven earlier in the context of the Ru-$L_{2,3}$ XAS spectra. Thus the spectrum of sample C [$V$(Ru)=4.22] also exhibits a small low-energy shoulder visible below the main Ru(IV) $t_{2g}$-feature. This shoulder is thus due to Ru(V) $t_{2g}$ hole states and is shifted to higher energy with respect to the $t_{2g}$-related feature in $Sr_2RuO_4$ in a manner fully analogous to the behaviour observed in the spectra of 3d-TM oxides due to a increase of covalence with increasing valence [13,28]. On further increasing the $V$(Ru) to 4.57 (sample J) the Ru(V) $t_{2g}$-feature becomes the dominant feature and the Ru(IV) $t_{2g}$-feature is no longer resolved as a separate peak. The observation of a spectral feature in the spectrum of $Sr_2Cu_{0.3}Ru_{0.7}O_4$ (sample J) at the same energy as the UHB in CuO (530.2 eV), indicates a similar Cu(II)-O bonding in this system, which is in agreement with the conclusions arrived at from the Cu-$L_3$ XAS spectrum.

The shift of the spectral features to lower energy as $V$(Ru) increases also means that for $Sr_4Ru_2O_9$ the $e_g$-related peak is shifted down and away from the Sr4d-related spectral weight, and thus is individually identifiable in the energy range 530.5-531.5 eV. It is evident that lower energy $t_{2g}$-related peak is narrower than the broad higher energy $e_g$-related peak. This is a direct result of the stronger covalent dpσ-like mixing between the O-2p and Ru-4d($e_g$) orbitals.

We now turn to the $La_{2-x}Sr_xCu_{1-y}Ru_yO_4$ compounds with $y \neq 0$ and $x \neq 0$ whose O-K spectra are shown in Fig. 8, together with the spectrum of $La_{1.8}Sr_{0.2}CuO_4$. The pre-edge peak in the latter is assigned to a hole state in the $CuO_6$ units induced by Sr doping [9,11,12]. In the Ru containing systems, if such a spectral feature were to occur it would overlap with the $t_{2g}$-related structure from Ru(V)$O_6$, which means that one cannot *a priori* distinguish between them. In fact, this hole state is only important in region (3) of the phase diagram (see Fig. 1), for example in sample P(1.36,0.2,0.28) where the $CuO_6$ units are doped with holes. In contrast to this, for the compounds in regions (1) and (2) of the phase diagram (Fig. 1), the O-K spectra should be a combination of the characteristic spectra from the Ru(IV)$O_6$, Ru(V)$O_6$, and Cu(II)$O_6$ units. Fig. 8 shows the O-K spectra of the selected compounds arranged with increasing $V$(Ru) from the bottom to the top. The spectrum of compound B(1.76,0.8,0) with $V$(Ru) = 4.2 shown at the bottom of Fig. 8 displays basically the same spectral features as $Sr_2RuO_4$, but with a lower-energy shoulder from a $t_{2g}$-related state in the Ru(V)$O_6$ unit due to the 20% Ru(V) ions present in sample B. With increasing Ru(V) content this state gains further in intensity and becomes the dominant peak at $V$(Ru) = 4.67 for sample K(1.68,0.6,0.04). In the case of sample L(1.5,0.5,0.05) with $V$(Ru) = 4.8 and N(1.28,0.4,0.04) with a Ru(V) state, only the $t_{2g}$-related state in the Ru(V)$O_6$ unit can be well recognized.

## IV. Summary

From a systematic study of $Cu\text{-}L_{2,3}$, $Ru\text{-}L_{2,3}$, and O-K XAS spectra of the system $La_{2-x}Sr_xCu_{1-y}Ru_yO_4$ we have found that for $x > 2y$, the charge balance upon Sr and Ru doping is achieved by an increase of the Ru valence from Ru(IV) to Ru(V) when going from region (1) to region (3) via region (2) of the phase diagram (Fig. 1), with Ru(V) being reached at the border between regions (2) and (3), while at all stages Cu stays essentially divalent. In region (3) of the phase diagramme, Ru remains as Ru(V), and charge balance forces a large oxygen deficit $\delta$ and a transfer of holes into the $CuO_6$ units, before leading finally to a breakdown of the $K_2NiF_4$-type structure at high $\delta$. Although hole counts of up to 0.2 holes per $CuO_6$ unit can be reached in region (3), these systems are not superconductors, which is most likely a consequence of the disturbance of the $CuO_2$ planes by the presence of the Ru ions. This effect is most probably primarily structural in nature (i.e. the continuous $CuO_2$ network is broken up), although the Ru ions can also act as magnetic pair breakers.

## V. Acknowledgements

We thank F. Grasset and J. Darriet for providing the $Sr_4Ru_2O_9$ sample and the staff of BESSY and HASYLAB for experimental assistance. This work was supported in part by the Deutsche Forshungsgemeinschaft within SFB 463 and SFB 484 and the Graduiertenkolleg 'Struktur- und Korrelationseffekte im Festkörper' at the Technical University Dresden.

## VII. References.

Figure captions

Fig. 1 Schematic phase diagram of $La_{2-x}Sr_xCu_{1-y}Ru_yO_4$, which is divided into three regions for $x > 2y$. The lines $x = 2y$, $x = 3y$ and $x = 2y + 0.4$ make region (1) [light grey shading], where we have the Ru valence $V(Ru) < 5$ and $\delta = 0$. The lines $x = 2y + 0.4$ and $y = 0.4$ make region (2) [dark grey shading], where we have $V(Ru) \leq 5$ and $\delta \geq 0$. The region (3) [hatched] lies below lines $x = 3y$ and $y = 0.4$, where we have $V(Ru) = 5$ and $\delta > 0$. The capital letters on the diagram indicate the stoichiometry of each sample, as listed in Tab. I. For detials, see text.

Fig. 2 Magnetic susceptibility, $\chi_{mol}$, of $La_2{}_{-x}Sr_xCu_1{}_{-y}Ru_yO_4{}_{-\delta}$, showing a Curie-Weiss behavior for P(1.36,0.2,0.28) (dashed line), a spin-glass behavior for E(1.64,0.7,0) (solid line), and an antiferromagnetic transition at $T_N \approx 117$ K for G(0.74,0.3,0) (dotted line).

Fig. 3 Cu-$L_{2,3}$ XAS spectra of selected $La_{2-x}Sr_xCu_{1-y}Ru_yO_4{}_{-\delta}$ compounds together with spectra of CuO and $La_2Li_{0.5}Cu_{0.5}O_4$ as Cu(II) and Cu(III) references, respectively.

Fig. 4 The XAS spectra of $Sr_2Cu_{1-y}Ru_yO_4$ and of $Sr_4Ru_2O_9$ as a Ru(V) reference at the Ru-$L_2$ edge (open circles) and at the Ru-$L_3$ edge (filled circles). The Ru-$L_2$ data have been shifted and multiplied to overlap with the Ru-$L_3$ data. The theoretical Ru-$L_{2,3}$ XAS spectra from crystal-field-multiplet calculations for Ru(IV) and Ru(V) are shown as solid and dashed lines stand for the $L_3$ and the $L_2$ edges, respectively. $V(Ru)$ indicates the Ru valence. For details, see text.

Fig. 5 Ru-$L_{2,3}$ XAS spectra of selected $La_{2-x}Sr_xCu_{1-y}Ru_y$ $O_4$ compounds together with that of $Sr_4Ru_2O_9$ as a Ru(V) reference. The $V(Ru)$ values indicate the Ru valence. The Ru-$L_2$ data have been shifted and multiplied to overlap with the Ru-$L_3$ data. For details, see text.

Fig. 6 (a) energy positions of the $t_{2g}$-related (filled circles) and the $e_g$-related (filled diamonds) features in the Ru-$L_3$ XAS spectra and (b) intensity ratios $I(t_{2g})/I(e_g)$ (filled squares) in the Ru-$L_2$ XAS spectra of $La_{2-x}Sr_xCu_{1-y}Ru_yO_4$ as a function of Ru valence. The open symbols show data for the Ru(V) reference compound $Sr_4Ru_2O_9$.

Fig. 7 O-K XAS spectra of $Sr_2Cu_{1-y}Ru_yO_4$, together with those of $Sr_4Ru_2O_9$ and CuO for comparison. In each case $V(Ru)$ indicates the Ru valence.

Fig. 8   O-K XAS spectra of selected $La_{2-x}Sr_xCu_{1-y}Ru_yO_4$. compounds, together with the spectrum of $La_{1.8}Sr_{0.2}CuO_4$ as a comparison. In each case $V$(Ru) indicates the Ru valence.

Table. 1   Ru valence $V$(Ru) of $La_{2-x}Sr_xCu_{1-y}Ru_yO_{4-\delta}$ as obtained from Eqs. (3)-(5) (see text). $\delta$ was obtained from thermogravimetric measurements.

| Sample name | X | Y | V(Ru) | d |
|:---:|:---:|:---:|:---:|:---:|
| A | 2 | 1 | 4 | 0 |
| B | 1.76 | 0.8 | 4.20 | 0 |
| C | 2 | 0.9 | 4.22 | 0 |
| D | 1.36 | 0.6 | 4.27 | 0 |
| E | 1.64 | 0.7 | 4.34 | 0 |
| F | 1.2 | 0.5 | 4.40 | 0 |
| G | 0.74 | 0.3 | 4.47 | 0 |
| H | 2 | 0.8 | 4.50 | 0 |
| I | 1.76 | 0.7 | 4.51 | 0 |
| J | 2 | 0.7 | 4.57 | 0.1 |
| K | 1.68 | 0.6 | 4.67 | 0.04 |
| L | 1.5 | 0.5 | 4.80 | 0.05 |
| M | 1.7 | 0.5 | 4.80 | 0.15 |
| N | 1.28 | 0.4 | 5 | 0.04 |
| O | 1.64 | 0.4 | 5 | 0.22 |
| P | 1.36 | 0.2 | >5 | 0.28 |

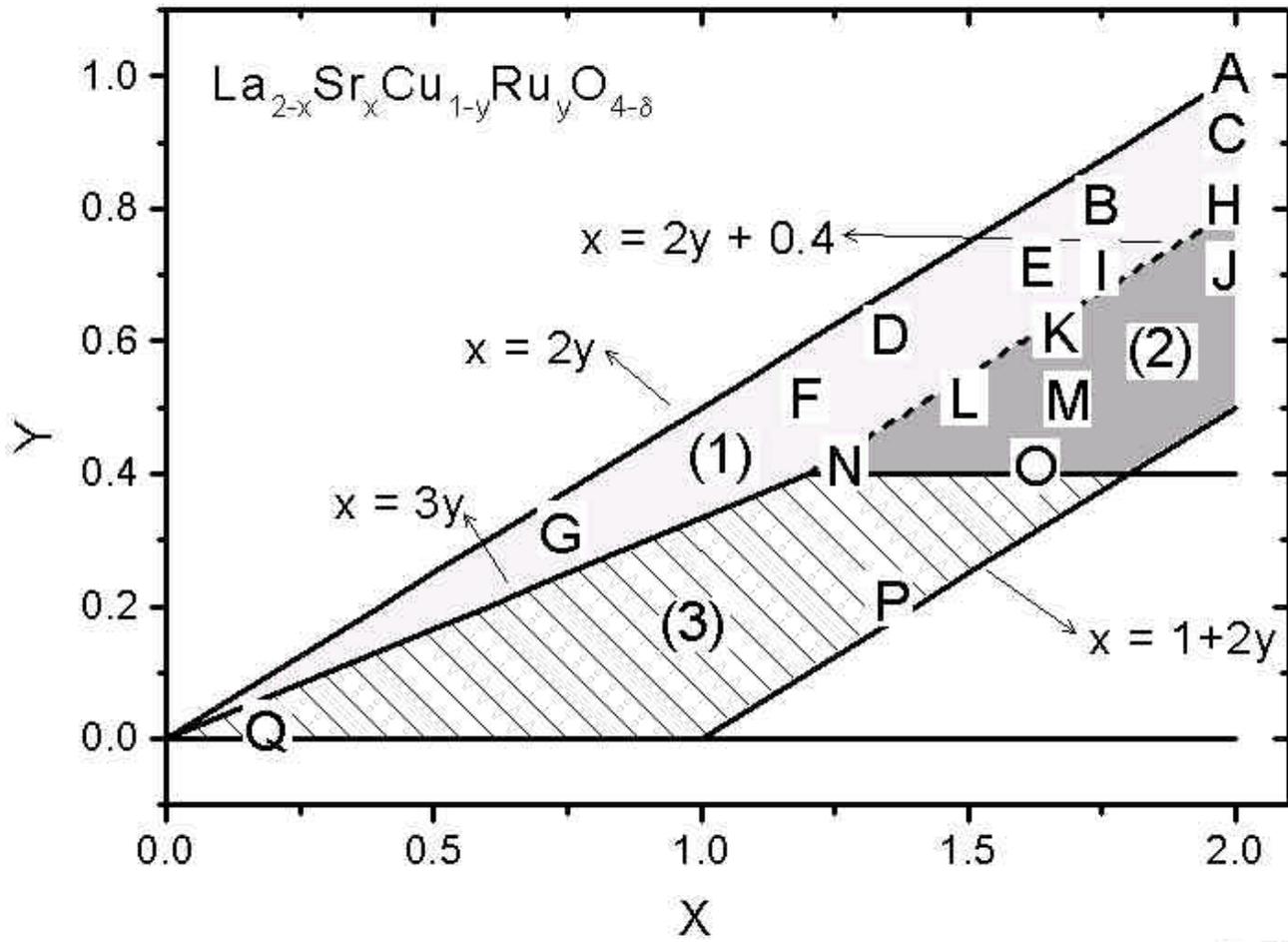

Z. Hu et al.
Fig. 1

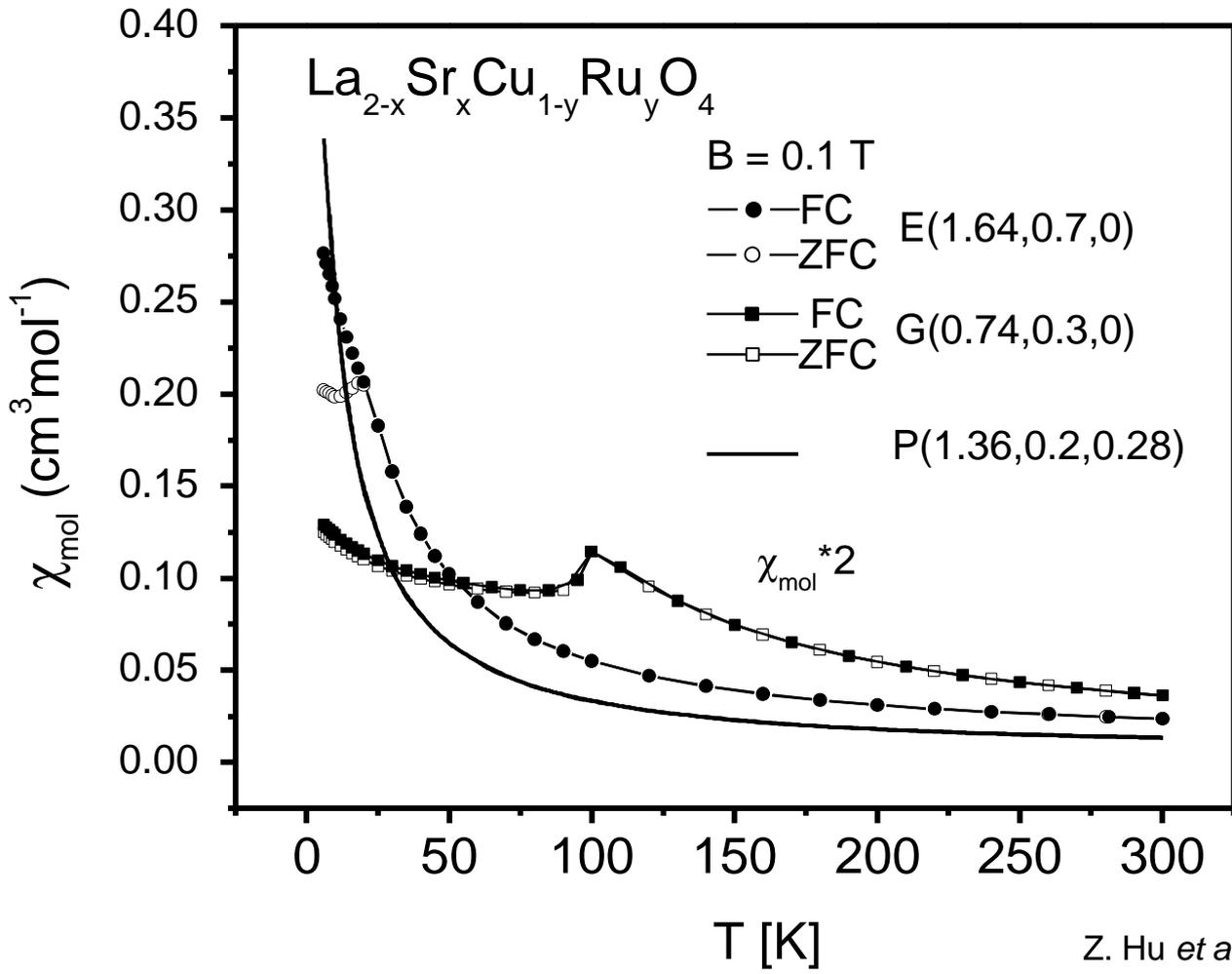

Z. Hu *et al.* Fig. 2

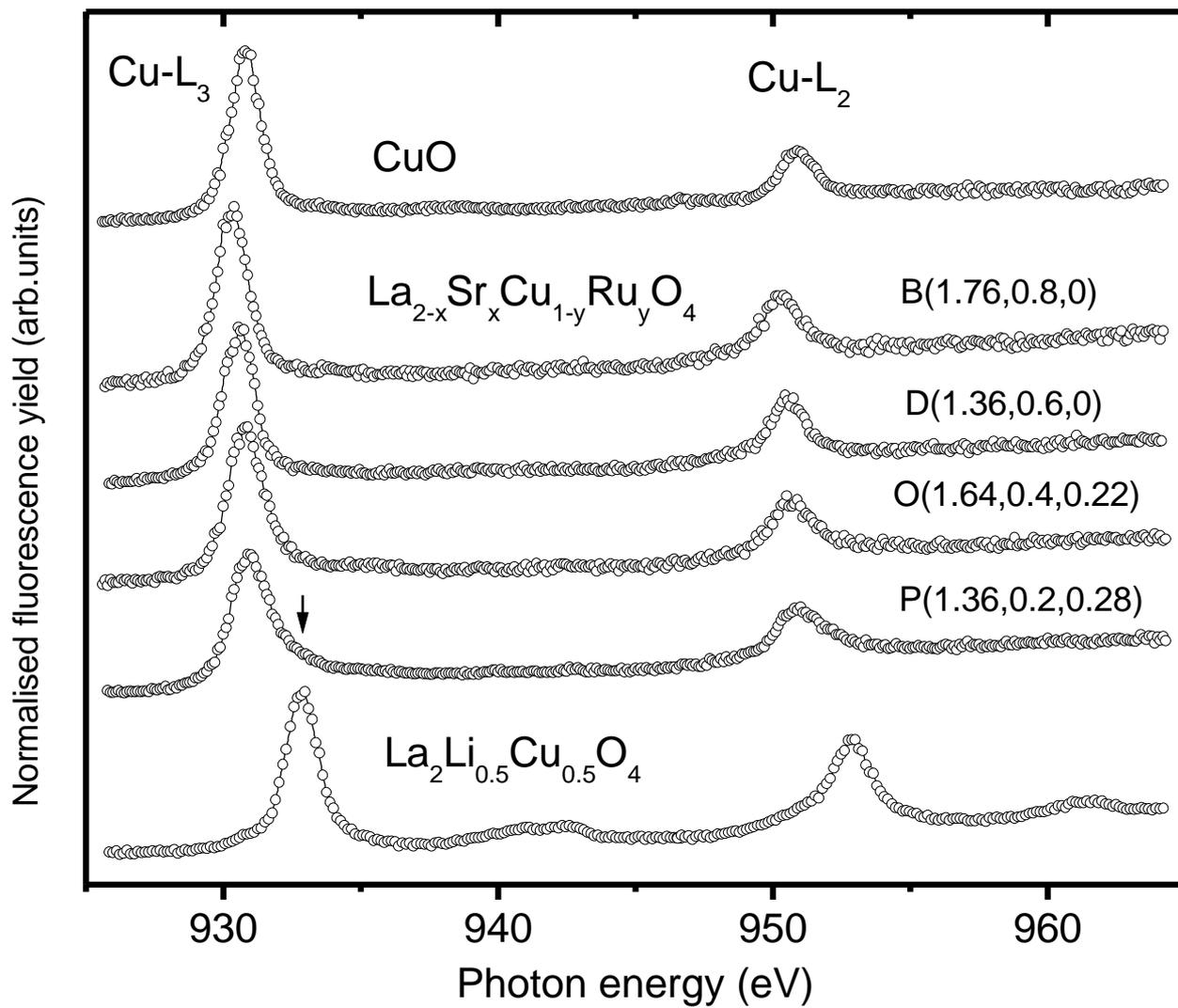

Z. Hu *et al.* Fig. 3

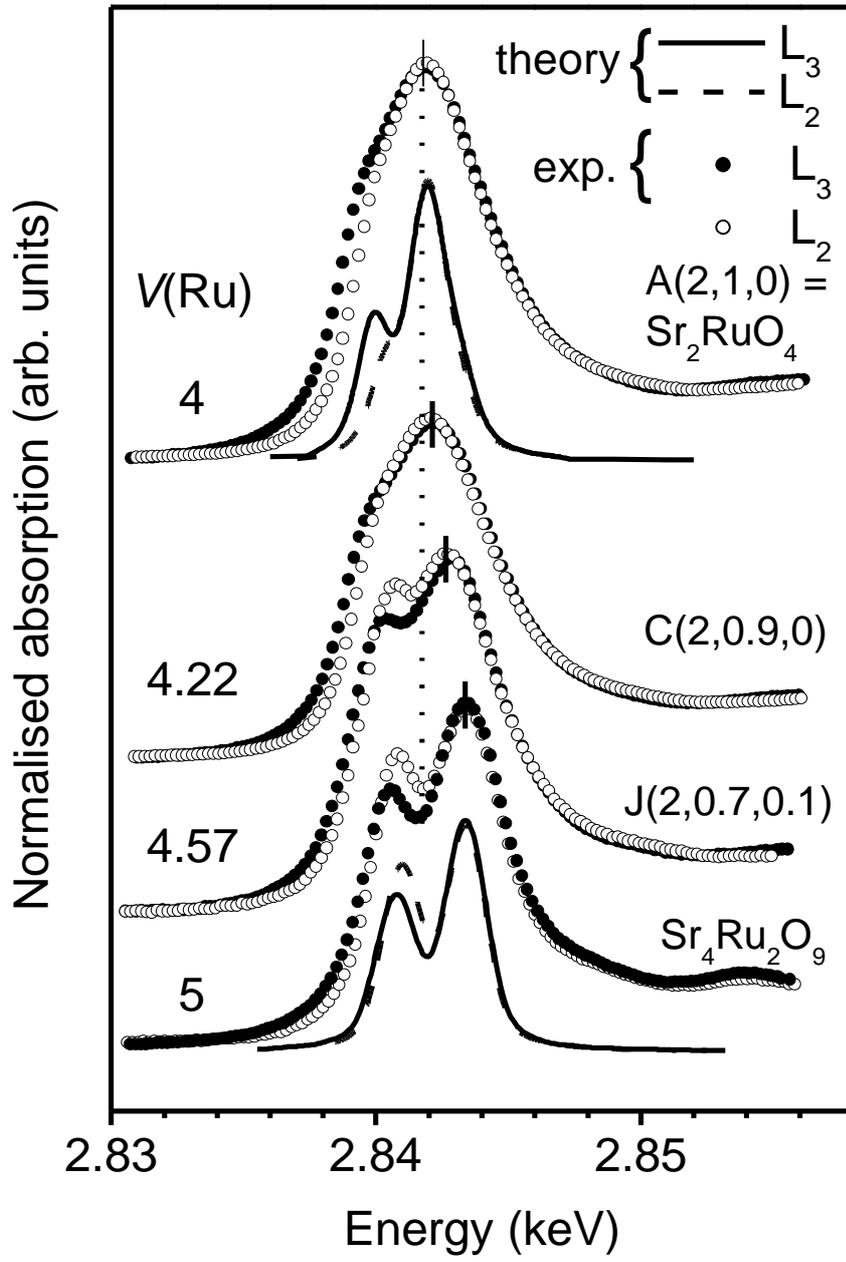

Z. Hu *et al.* Fig. 4

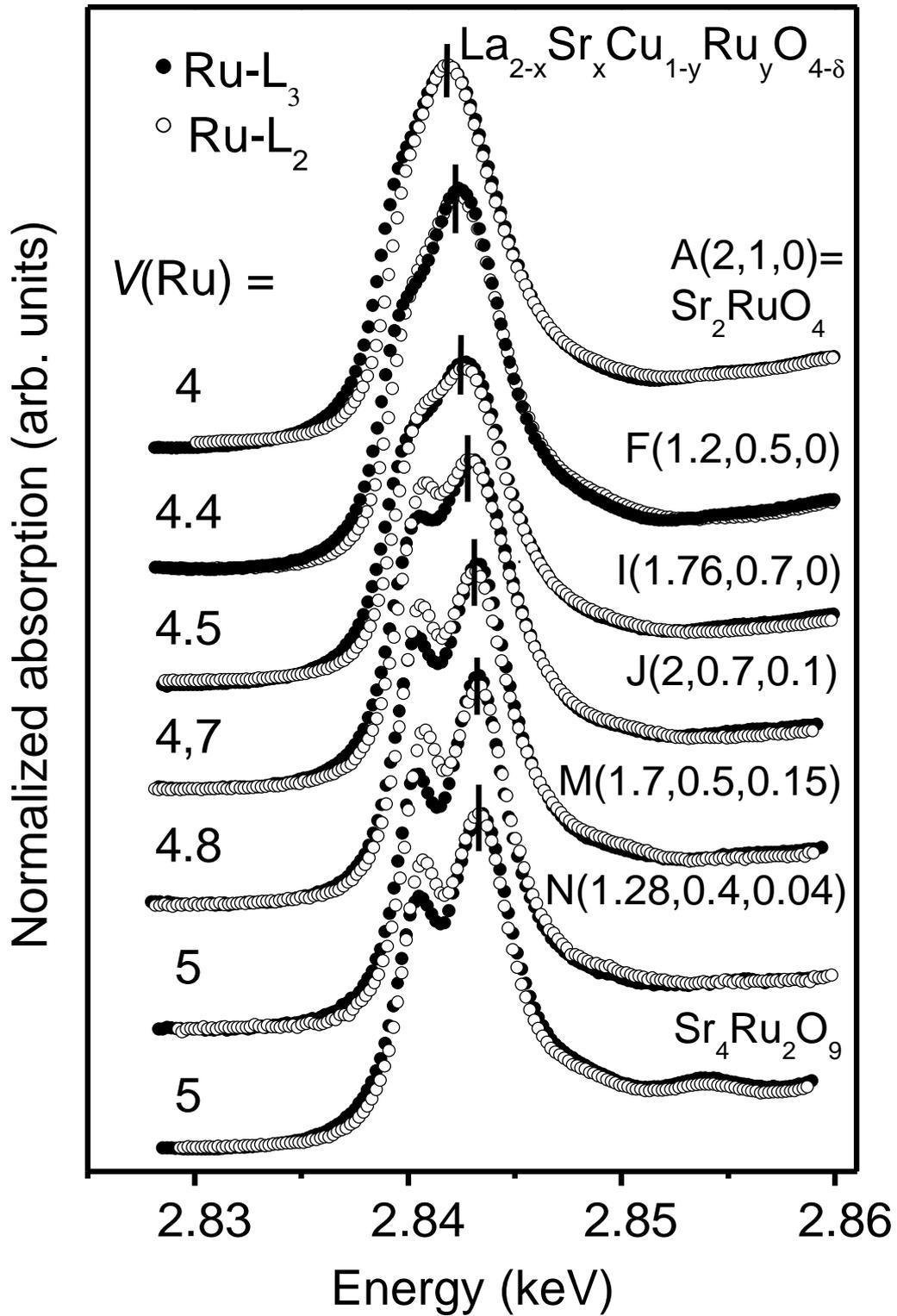

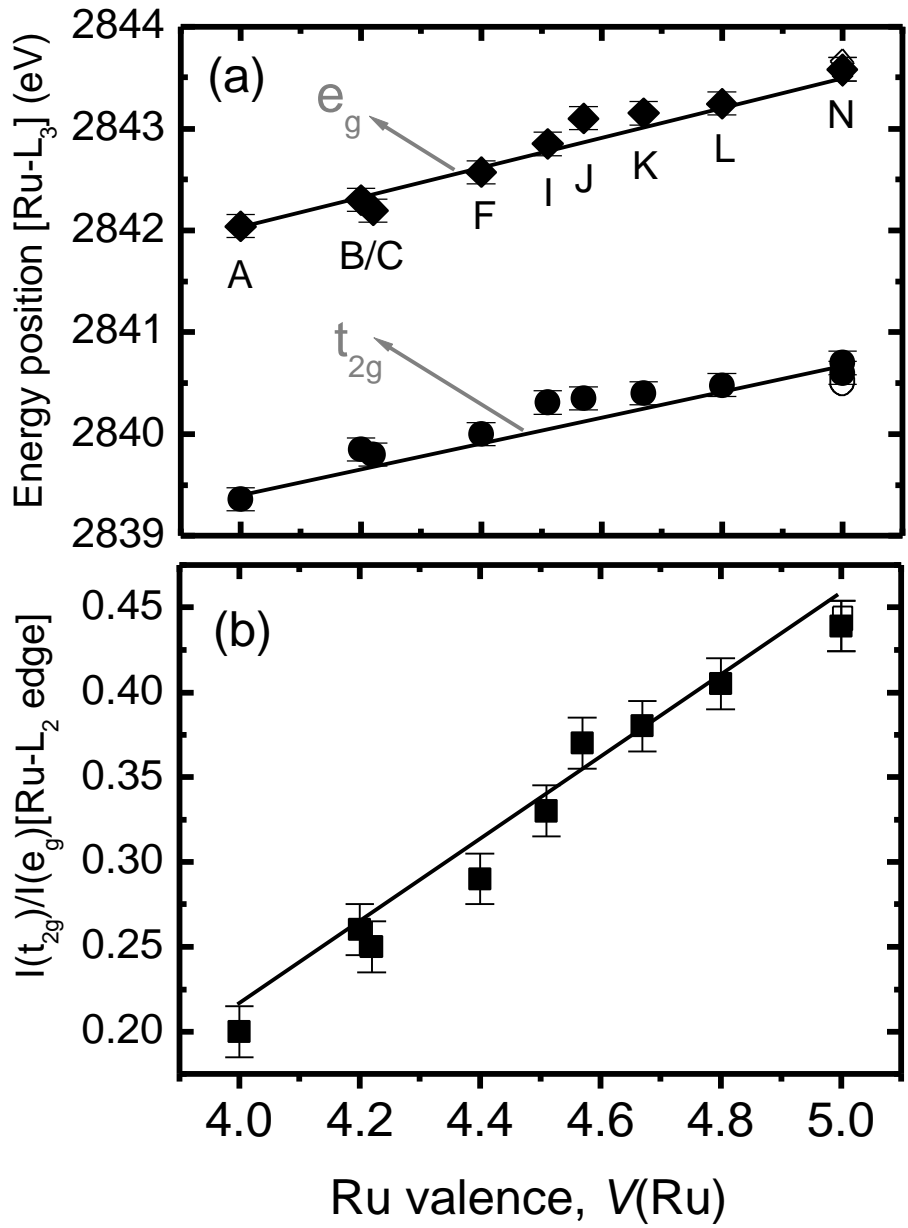

Hu *et al.* Fig. 6

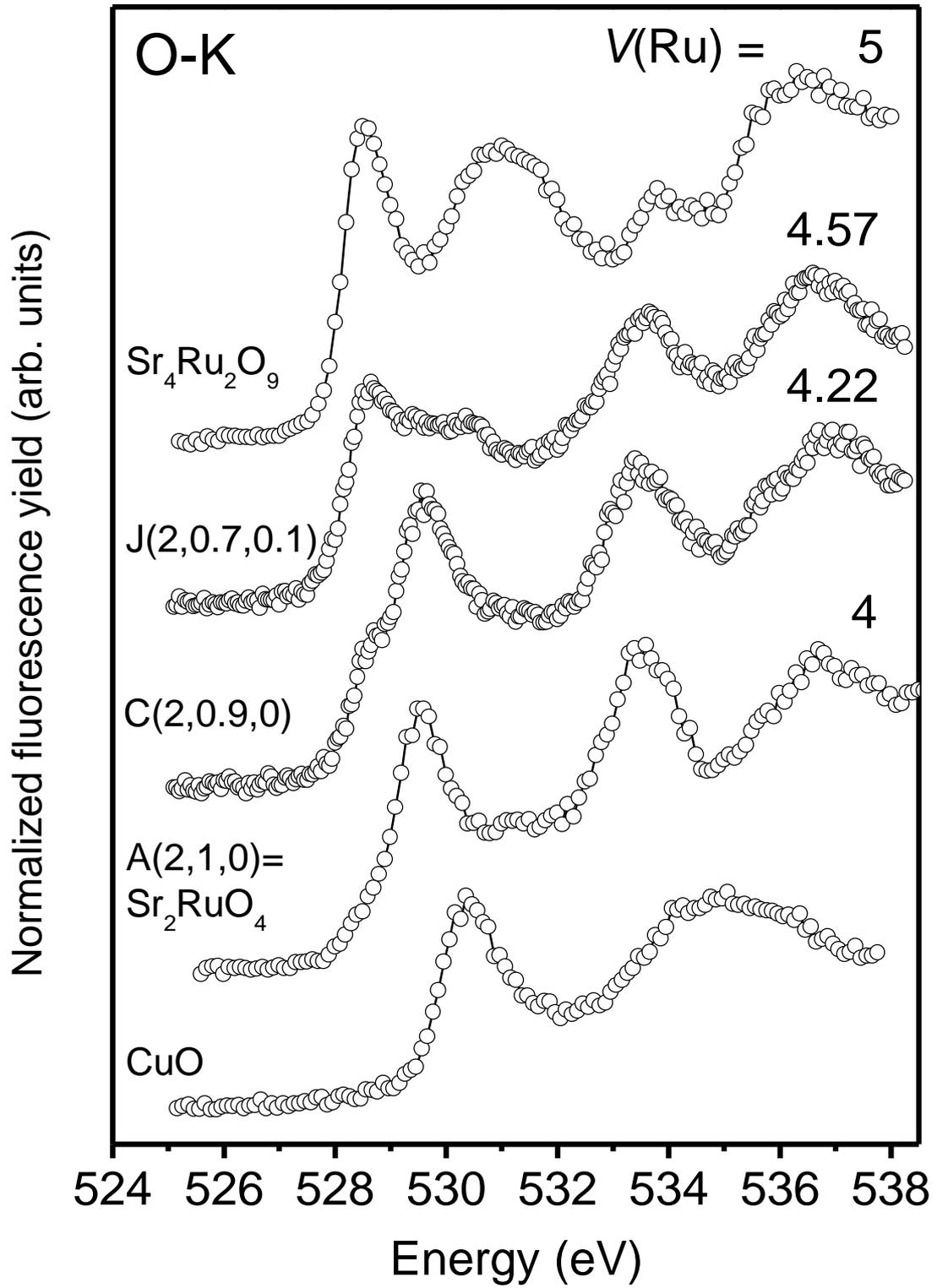

O-K

$V$(Ru) = 5

4.57

4.22

$Sr_4Ru_2O_9$

J(2,0.7,0.1)

C(2,0.9,0)

A(2,1,0)=
$Sr_2RuO_4$

CuO

Normalized fluorescence yield (arb. units)

Energy (eV)

Hu *et al.* Fig. 7

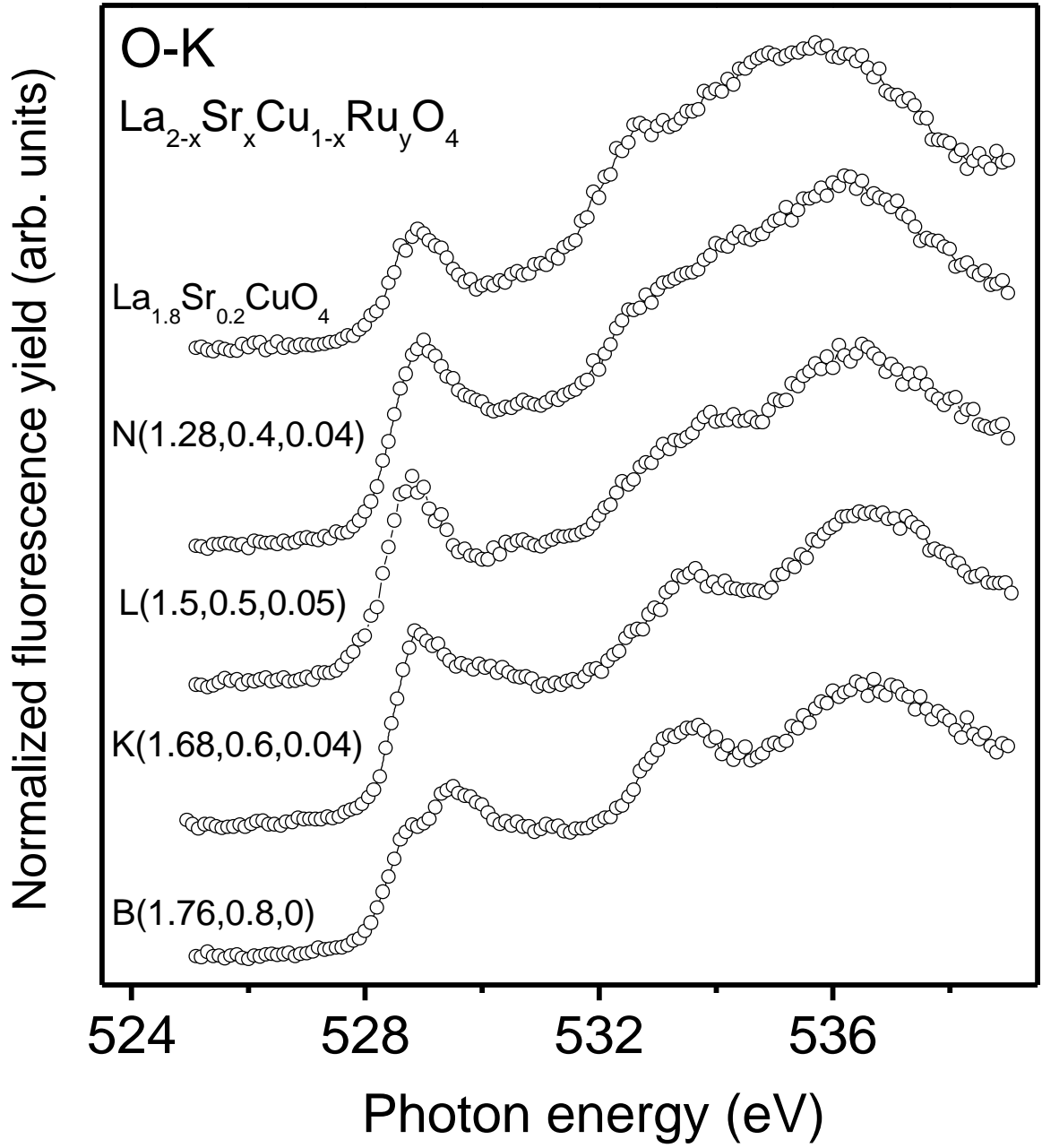

O-K

$La_{2-x}Sr_xCu_{1-x}Ru_yO_4$

$La_{1.8}Sr_{0.2}CuO_4$

N(1.28,0.4,0.04)

L(1.5,0.5,0.05)

K(1.68,0.6,0.04)

B(1.76,0.8,0)

Normalized fluorescence yield (arb. units)

Photon energy (eV)

Hu *et al.* Fig. 8